\documentclass[3p,times]{elsarticle}

\usepackage{ecrc}

\volume{00}

\firstpage{1}

\journalname{Nuclear Physics A}

\runauth{}

\jid{nupha}

\usepackage{amsmath}

\usepackage{amssymb}
\usepackage[figuresright]{rotating}

\newcommand \bea{\begin{eqnarray} }
\newcommand \eea{\end{eqnarray} }

\begin{document}

\begin{frontmatter}



\dochead{}

\title{Calculating the Jet Transport Coefficient $\hat{q}$ in Lattice Gauge Theory }


\author{Abhijit Majumder}

\address{Department of Physics and Astronomy, Wayne State University, Detroit, Michigan, 48201, USA.}

\begin{abstract}
The formalism of jet modification in the higher twist approach is modified to describe a hard parton propagating through a 
hot thermalized medium.
%We present a framework where first principles calculations of jet modication may be carried out in a non-perturbative thermal environment. As an example of this approach, we compute 
The leading order contribution to the transverse momentum broadening of a high energy (near on-shell) quark in a thermal medium is 
calculated. This involves a factorization of the perturbative process of scattering of the quark from the non-perturbative transport coefficient. An operator product expansion of the non-perturbative operator product which represents $\hat{q}$ is carried out and related via dispersion relations to the expectation of local operators. These local operators are then evaluated in quenched SU(2) lattice gauge theory.
\end{abstract}


\end{frontmatter}

% \linenumbers


\section{Introduction}
\label{intro}

Jet quenching, the modification of hard jets~\cite{Gyulassy:1993hr} in a hot dense extended medium, is considered as one of the 
most penetrating probes of the Quark Gluon Plasma (QGP) formed in high energy heavy-ion collisions. 
At this time, the most promising approach to jet modification has been to treat the propagation of the hard patrons in the jet using 
perturbative QCD (pQCD), factorized from the soft medium which is treated non-perturbatively~\cite{Majumder:2010qh}.
In so doing, the effect of the medium has been codified into a handful of transport coefficients: The transverse momentum squared gained per unit length, by a single parton, without radiation ($\hat{q} = \langle  p_{\perp}^{2} \rangle/t$),  
the longitudinal momentum lost per unit length without radiation ($\hat{e} = \langle  p_{L} \rangle/t $), 
the fluctuation in longitudinal momentum per unit length ($\hat{e}_{2} = \langle  p_{L}^{2} \rangle/t $)~\cite{Majumder:2007hx,Majumder:2008zg,Muller:2012hr}.

As derived in the references mentioned above, all of these transport coefficients have well defined operator 
expressions in a given gauge and also have gauge invariant definitions derived in effective field theory~\cite{Idilbi:2008vm,Benzke:2012sz}.
However, to this day, there is no first principles calculation of these coefficients within QCD at the temperatures of relevance at RHIC and 
LHC collisions. In these proceedings, we report on such an attempt~\cite{Majumder:2012sh}.

\section{A single parton in a QGP brick}
\label{formal}

In this attempt, the simplest example of jet modification in a finite thermalized medium will be considered: A single quark propagating through a box of length $L$, held at a fixed temperature, undergoing a single scattering which imparts transverse momentum to the quark. 
The quark then exits the medium and is observed. The length of the medium will be considered to be short enough such that the 
quark does not radiate during traversal. The coupling of the jet with the medium will be considered to be small enough such that secondary scattering is minimal. 
%The transverse momentum coefficient $\hat{q}$, will be derived from this single scattering cross section. 

Consider a quark in a well defined momentum state $ | q \rangle ( \equiv |q^+, q^- ,0_\perp \rangle )$ 
impinging on a medium $| n \rangle $ and then exiting with transverse momentum $k_{\perp}$, 
\bea
 | q + k \rangle \equiv \left| \frac{\left( k_\perp^2 + Q^2 \right)}{[2 (q^- ) ] } , q^- ,  \vec{k}_\perp \right\rangle.
\eea
The medium state absorbs this change in momentum and becomes $| X \rangle$. The quark is assumed to 
be space-like off-shell with virtuality $Q^2 = 2 q^+q^- \leq 0$ with the negative $z$-axis defined as the direction of the 
propagating quark. 
The spin color averaged transition probability (or matrix element) for this 
process, in the interaction picture,  is given as 
\bea
W(k) =  \sum_{X} \frac{1}{2N_c } \frac{\sum_{n}}{Z} e^{-\beta E_{n}}\langle q^- ; n |  T^* e^{ i \int_0^{t} d  t H_I(t)  } | q^-+k_\perp , X \rangle
 \langle   q^-+k_\perp , X  | T e^{-i \int_0^{t} dt H_I(t) }  | q^- ,n \rangle ,
\eea  
where, we have averaged over the initial color and spin of the quark. Assuming that the medium is in a thermalized state we 
have also averaged over the various initial states of the medium, weighted with a Boltzmann factor ($Z$ represents the partition 
function of the medium without the incoming quark). 
Summing over the various values of $k_{\perp}$, weighted with $k_{\perp}^{2}$, and dividing by the length yields the transverse 
momentum diffusion coefficient,  
\bea
\hat{q} = \sum_{k}  k_\perp^2  \frac{W(k)}{t}.
\eea

Expanding the time evolution operator to leading order in coupling, converting factors of 
$k_{\perp}$ into transverse gradients (and finally into field strength operators), and summing over the final state $|X\rangle$, we obtain the non-perturbative definition of the transport 
coefficient $\hat{q}$, i.e., 
\bea
\hat{q} = \frac{4 \pi^2 \alpha_s}{ N_c } \int \frac{dy^- d^{2} y_{\perp}}{(2 \pi)^{3}} d^{2} k_{\perp} 
e^{ -i \frac{k_{\perp}^{2}}{2q^{-} }  y^{-} +  i\vec{k}_{\perp} \cdot \vec{y_{\perp} } } 
\langle n |\frac{e^{-\beta E_n}}{Z} {F^{+,}}_\perp (y^-) F_\perp^+ (0)  | n \rangle.
\eea
In the above expression, there is no ordering between the two field strength operators. 
The expression above is not gauge invariant, but is gauge covariant.  As a result, if one were to carry out an 
operator product expansion in terms of local operators, one could reorganize the expansion to only contain 
gauge invariant local operators. Any gauge dependence would then only be contained in the coefficient functions. 
The operators in the product above are, however, separated by a light like distance. As a result, a straightforward 
expansion in $y^{-}$ will not suffice.

\section{Analytic Continuation and Finite Temperature Dispersion Relations }
\label{anal}

The expression for $\hat{q}$ above may be written as the discontinuity of a more generalized operator product, 
\bea
\hat{q} = {\rm Disc} \left[ \hat{Q} \right] &=& {\rm Disc} \left[  \frac{4 \pi^{2} \alpha_{s}}{N_{c}}\!\!\! \int \frac{d^{4}y d^{4} k}{(2\pi)^{4}} 
e^{i k \cdot y} \frac{2 (q^{-})^{2}}{ \sqrt{2} q^{-} } 
\frac{ \sum_{n} e^{-\beta E_{n} } }{Z} \frac{ \langle n | F^{+ \perp}(0) 
F_{\perp,}^{+}(y) | n \rangle}{ (q+k)^{2} + i \epsilon }  \right] .
\eea
The generalized quantity $\hat{Q}$ should be considered as an analytic function of $q^{+} , q^{-}$. Keeping $q^{-}$ constant 
as the large scale $Q$ in the problem, we consider $\hat{Q}$ as a function of $q^{+}$. The jet transport coefficient is the discontinuity 
in $\hat{Q}$ in the region where $|q^{+}| \ll | q^{-} | $.

We now consider the integral, 
\bea
I_{1} = \oint \frac{d q^{+}}{2 \pi i} \frac{  \hat{Q}(q^{+}) }{ \left( q^{+}  + Q_{0} \right) },
\eea
where $Q_{0}$ is positive and of the order for . The contour is taken as a small counter-clockwise circle 
around the point $q^{+} = - Q_{0}$. The residue of this integral is given as, 
\bea
I_{1} = \left. \hat{Q}(q^{+}) \right|_{q^{+} = - Q_{0}} .
\eea
When $q^{+} = Q_{0} \sim Q$, the hard scale in the problem, one can expand the denominator in $\hat{Q}$ as 
a series of local operators involving ever higher powers of derivatives, 
\bea
\hat{Q} = \frac{4\sqrt{2} \pi^{2} \alpha_{s} q^{-}}{N_{c}  Q^{2}} 
 \sum_{n} \frac{e^{-\beta E_{n}}}{ Z}  \langle n | F^{+ \mu}_{\perp}
 \sum_{m=0}^{\infty} \left( \frac{ -q \cdot i\mathcal{D} - \mathcal{D}_{\perp}^{2} }{Q^{2}} \right)^{m} 
 F^{+}_{\perp, \mu}  | n \rangle. \label{expansion}
 \eea
We can now deform the contour to have it encircle the cut in the region where $| q^{+} | \ll Q$ the hard scale. 
The discontinuity in this region, is related to the discontinuity of $\hat{Q}$ which in turn leads to the physical transport 
coefficient $\hat{q}$. In this region, however, one cannot use the expansion of Eq.~\eqref{expansion}. Instead one obtains, 
\bea
I_{1} = \frac{4 \pi^{2} \alpha_{S}}{N_{c}}  \int d q^{+} \frac{d^{4} y d^{4} k}{ (2\pi)^{4} } e^{ik \cdot y} 
\frac{\delta \left( k^{+} + q^{+} - \frac{k_{\perp}^{2}}{ 2 q^{-}} \right) }{ 2 q^{-} }
\sum_{n} \frac{e^{-\beta E_{n}}}{Z} \frac{ \langle n | F^{+ \mu} (0) F^{+}_{\mu , }(y)  | n \rangle }{ \left( q^{+} + Q_{0} \right)  } 
= \int_{-\lambda^{2} Q}^{\lambda^{2} Q} d q^{+}  \frac{\hat{q} (q^{+})}{ q^{+} + Q_{0}} + \int_{0}^{\infty} d q^{+} V(q^{+}). 
\label{physical-expansion}
\eea
The second integral in the equation above, represents the vacuum process of a time-like quark decaying by the emission of 
a hard gluon. 
The dimensionless quantity $\lambda$ in the limits of integration above is meant to indicate a small number $\lambda \ll 1$. 
It merely states that the physical $\hat{q}$ integral is over a small range in $q^{+}$, far from $q^{+} = Q_{0}$.
For 
a jet with maximum virtuality $\mu^{2}$ and $(-)$ momentum $q^{-}$, $\lambda^{2} Q = \mu^{2}/(2q^{-})$.
Given this small range of values of $q^{+}$, the integrand can now be expanded as a Taylor expansion in $q^{+}/Q_{0}$.
This series can then be compared with the series of Eq.~\eqref{expansion}, and terms equated to obtain the various 
moments of $\hat{q}$.

The methodology outlined above can be made even more precise and straightforward by setting a definite value for 
$Q^{0} = q^{-}$. While this will readjust the relative importance of the various terms in the series it allows for simpler 
set of operators that need to be evaluated numerically.  This simplifies $I_{1}$ in Eq.~\eqref{expansion} to, 
\bea
I_{1} &=&  \frac{ 2 \sqrt{2} \, \pi^{2} \alpha_{s}}{N_{c} \, q^{-} } \sum_{n} \frac{e^{-\beta E_{n}}}{ Z }  \langle n | \, F^{+ \mu}_{\perp}
 \sum_{n=0}^{\infty} \left( \frac{ -i\mathcal{D}^{0} }{q^{-}}  \right)^{n}  F^{+}_{\perp, \mu} \, | n \rangle,  \label{I1-simple}
\eea
and similarly simplifies Eq.~\eqref{physical-expansion} with $Q_{0}$ replaced by $q^{-}$. For a virtuality $\mu^{2}$ such that 
$\Lambda_{QCD}^{2} \ll \mu^{2} \ll (q^{-})^{2}$, we can define a $q^{+}$ or virtuality averaged $\hat{q}$ as, 
\bea
\hat{\bar{q}} (Q^{+}) 2Q^{+}  = \int_{-Q^{+}}^{Q^{+}} dq^{+} \hat{q}(q^{+}) 
\simeq 2 \hat{q} Q^{+} + \frac{ \hat{q}'' (Q^{+})^{3} }{3},
\label{mean-qhat}
\eea 
where the last partial equality is only valid in the limit that $\hat{q}$ is a slow function of $q^{+}$,
%(or alternatively stated $Q^{+} \ll q^{-} $) 
note $Q^{+} = \mu^{2}/q^{-}$. One can now compare 
coefficients of $q^{-}$ between Eq.~\eqref{I1-simple} and Eq.~\eqref{physical-expansion}, using the mean 
values defined in Eq.~\eqref{mean-qhat}.

\section{Lattice calculation and Results}
\label{res}

The remaining methodology consists of simply evaluating the series of operator products in Eq.~\eqref{I1-simple} 
on an lattice at finite temperature and then comparing them with terms with the same power of $q^{-}$ in 
Eq.~\eqref{physical-expansion}. In order to calculate the thermal expectation values of the various local operators 
we have to simply rotate the time coordinate to the imaginary time direction: $x^{0} \rightarrow i x^{4}$, $A^{0} \rightarrow iA^{4}$
and $F^{0 i} \rightarrow i F^{4i}$.

For this exploratory study we calculate in quenched $SU(2)$ gauge theory, as 
a simpler substitute for QCD. Calculations are carried out on a lattice of spatial size $n_{S} = n_{T} \times 4$, with temporal 
extent $n_{T}$ varying from 3 to 6. The scale is set using the renormalization group formula~\cite{Engels:1980ty,Creutz:1984mg}, 
\bea
a_{L} = \frac{1}{\Lambda_{L}} \left(\frac{11 g^{2}}{ 24 \pi^{2}}\right)^{-\frac{51}{121}} \exp \left( - \frac{12\pi^{2} }{11 g^{2}} \right). \label{lattice-spacing}
\eea
In the equation above, $a_{L}$ is the lattice spacing, $g$ represents the bare lattice coupling and $\Lambda_{L}$ represents the one 
dimension-full parameter on the lattice. Comparing with the vacuum string tension, we have used $\Lambda_{L}=5.3$~MeV.   
For a lattice at finite temperature or one with $n_{t} \ll n_{s}$, the temperature is obtained as
\bea
T = \frac{1}{n_{t} a_{L}}.
\eea

\begin{figure}[h!]
%\begin{center}
%  \epsfxsize 80mm
%\hspace{0cm}
\hspace{0.25in}
\resizebox{2.65in}{2.65in}{\includegraphics{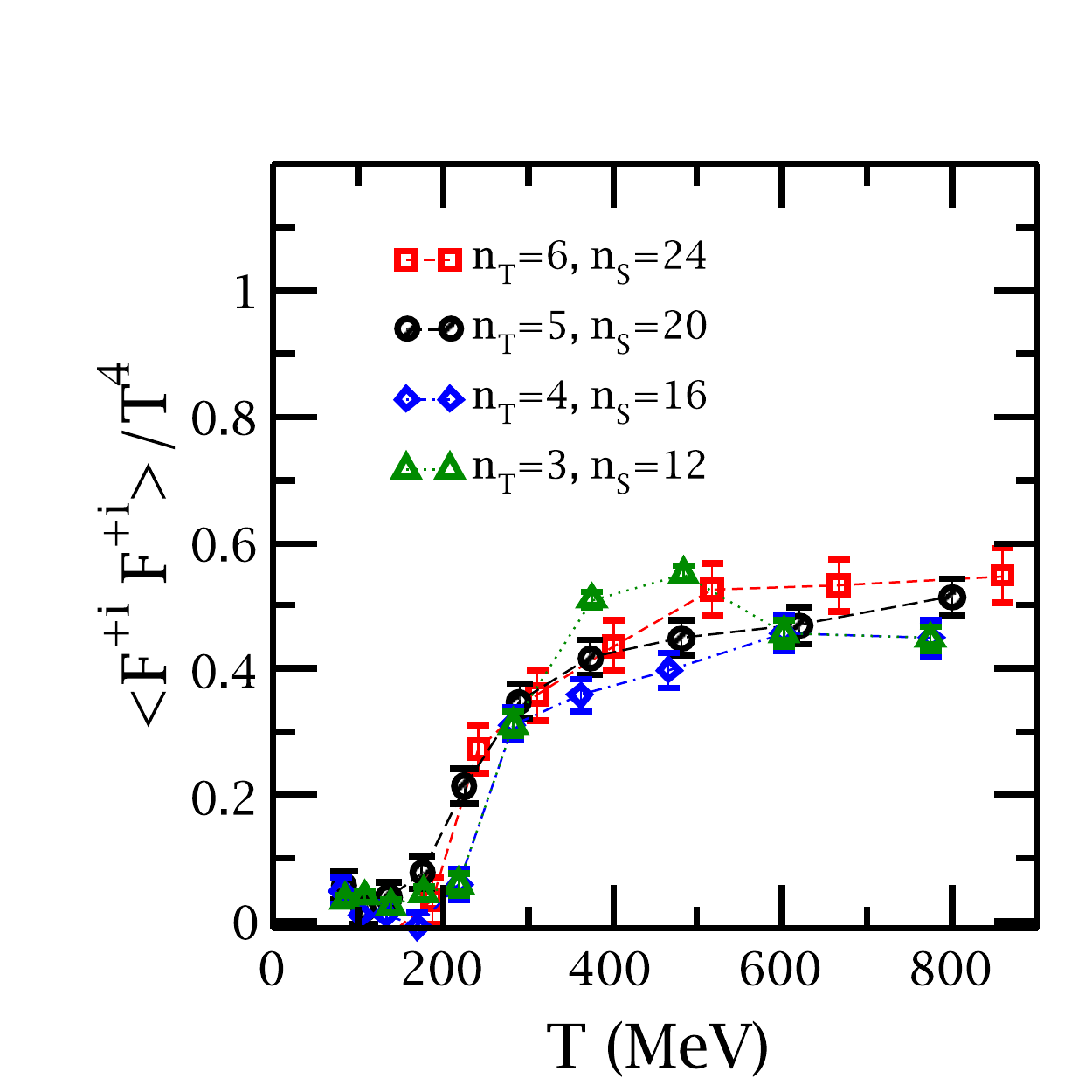}} 
\hspace{0.5in}
\resizebox{2.65in}{2.65in}{\includegraphics{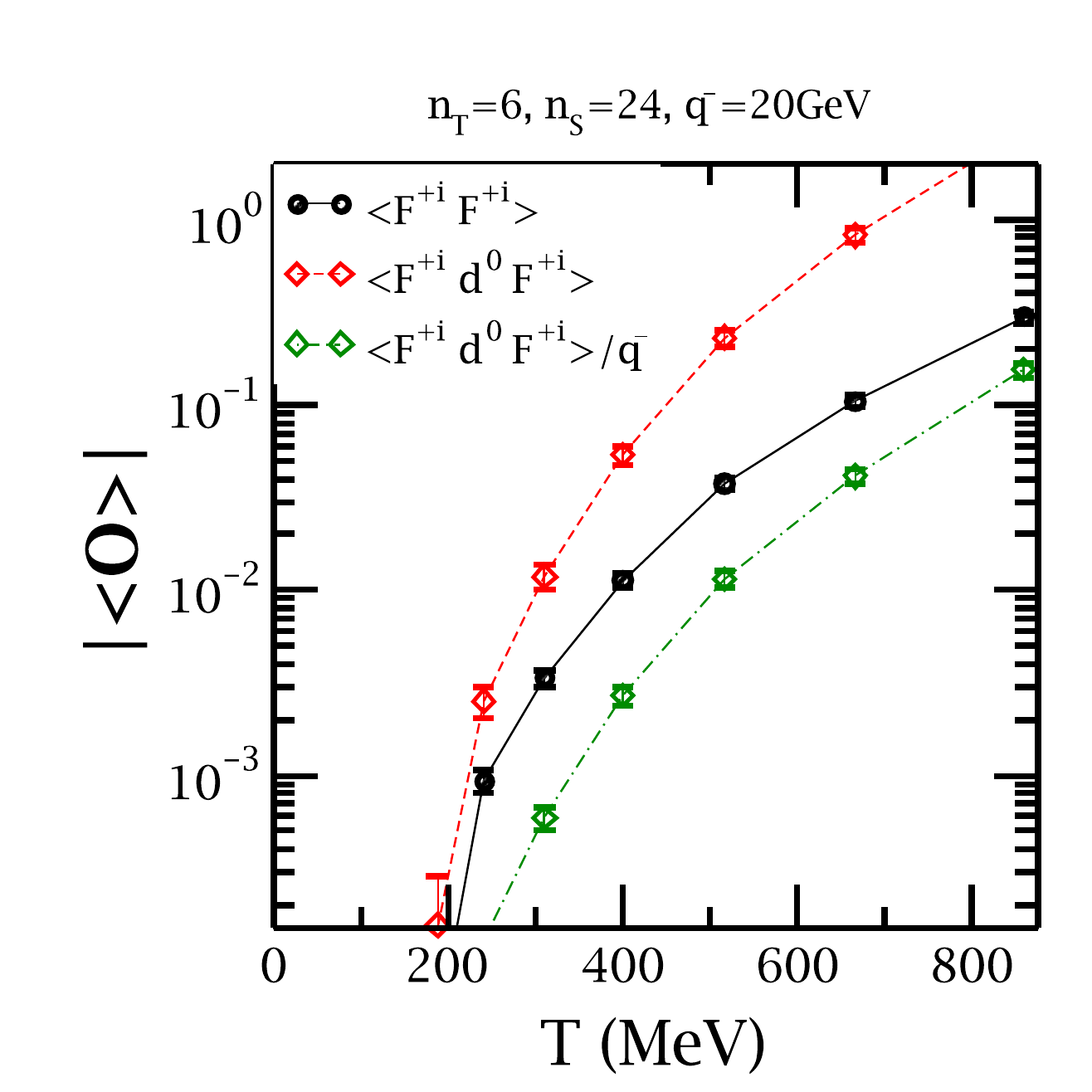}} 
%\vspace{0.25cm}
    \caption{Left panel: The temperature dependence of the local operator $\langle F^{+ i} F^{+ i} \rangle$ scaled by $T^{4}$ to make it 
    dimensionless. Right panel: temperature dependence of absolute values of the the local operator $\langle F^{+ i} F^{+ i} \rangle$ and the next-to-leading operator products $\langle \left[ F^{+ i} i \mathcal{D}^{4} F^{+ i} \right] \rangle$ [red diamond (dashed)] and 
    $\langle \left[ F^{+ i} i \mathcal{D}^{4} F^{+ i} \right]/q^{-} \rangle$ [green diamond (dot dashed)].}
    \label{fig1}
%  \end{center}
\end{figure}

The results reported in this proceedings are the result of 5000 heat bath sweeps per point. 
In the left panel of Fig.~\ref{fig1} we present the results of the lattice calculation for the first operator 
product in Eq.~\eqref{I1-simple}, i.e., $\langle F^{+ i} F^{+ i} \rangle$ scaled by $T^{4}$ to make it 
dimensionless.
With these statistics, we find acceptable 
scaling with lattice size, especially in the region above and away from the phase transition, i.e., for $T>400$~MeV.
In the right panel we calculate the second term  in the expansion, $\langle F^{+ i} \mathcal{D}^{0} F^{+ i} \rangle$, in Eq.~\eqref{I1-simple}, both with (green diamonds) and without 
(red diamonds) the large factor of $q^{-}$ in the denominator. While this operator product is by itself larger than the leading 
term, the large factor of $q^{-}$ in the denominator makes it considerably smaller than the leading term in the region with 
$T < 600$~MeV. As a result, in the region with $400$~MeV$ < T < 600$~MeV, the leading order term in Eq.~\eqref{I1-simple} 
will yield an usable estimate for  $\hat{q}$.

To estimate $\hat{q}$ we use the point at $T=400$~MeV, where $\langle F^{+ i} F^{+ i} \rangle = 0.01$~GeV$^{4}$.
We are considering a lattice with a length given by $4 \times n_{t} a_{L} = 4/0.4$~GeV$^{-1} = 10$~GeV$^{-1}$. 
This states that the maximum virtuality of a jet (with a $q^{-} = 20$~GeV) which traverses such a length without undergoing 
radiation is given as $\mu^{2} = E/L = 20/10/\sqrt{2} \simeq 1.4$~GeV$^{2}$. Thus $Q^{+}=1.4/40$~GeV. We  can now use the 
 formula  
$
\hat{\bar{q}} = \frac{2 \sqrt{2} \pi^{2} \alpha_{s}(\mu^{2})  }{N_{c} 2 Q^{+} q^{-} }  \langle F^{+ i}  F^{+ i}  \rangle.
$
Using $\alpha_{s} (1.4 \rm{GeV}^{2}) = 0.375$~\cite{Kluth:2007np}, we obtain $\hat{\bar{q}} = 0.134 \rm{GeV}^{2}/{\rm fm}$ for an 
$SU(2)$ quark traversing a quenched $SU(2)$ plasma. In most phenomenological estimates one quotes the $\hat{q}$ of the 
gluon. 
If the above calculation were done for an $SU(2)$ gluon, the $\hat{q}$ would 
differ only by the overall Casimir factor of $C_{A}/C_{F} = 2N_{c}^{2}/(N_{c}^{2} - 1) = 8/3$ yielding a $\hat{q}_{G} = 0.358$~GeV$^{2}$/fm, 
at a $T \sim 400$~MeV (Note, this value is different from that in Ref.~\cite{Majumder:2012sh} by a factor of 4, due to 
the neglect of this factor in that reference).



%% References


\bibliographystyle{elsarticle-num}
\bibliography{refs}






% \end{thebibliography}

\end{document}